\documentclass[draft, sort&compress, draft]{aipproc}

\layoutstyle{8x11double}
\usepackage{natbib}
\usepackage{graphicx}

\begin{document}
\title{Invariance of Structure in an Aging Colloidal Glass}
\author{Gianguido C. Cianci}{address=
{Department of Physics, Emory University, Atlanta, GA 30322, U.S.A.}}
\author{Rachel E. Courtland}{address=
{Department of Physics, Emory University, Atlanta, GA 30322, U.S.A.}}
\author{Eric R. Weeks}{address=
{Department of Physics, Emory University, Atlanta, GA 30322, U.S.A.}}

\begin{abstract}
We study concentrated colloidal suspensions, a model system
which has a glass transition.  The non-equilibrium nature
of the glassy state is most clearly highlighted by aging --
the dependence of the system's properties on the time elapsed
since vitrification. Fast laser scanning confocal microscopy
allows us to image a colloidal glass and track the particles
in three dimensions. We analyze the static structure in terms
of tetrahedral packing. We find that while the aging of the
suspension clearly affects its dynamics, none of the geometrical
quantities associated with tetrahedra change with age.
\end{abstract}

\maketitle


\section{Introduction}

If a liquid is rapidly cooled faster than it can crystallize, it
can form an amorphous solid also known as a glass.  This occurs
below a temperature known as the glass transition temperature,
$T_{\rm g}$.  As $T_{\rm g}$ is approached from above, the viscosity
of the liquid increases by many orders of magnitude over a
small range in temperature.  However,
a diverging structural length scale that would also grow by
orders of magnitude and thus explain this thickening has not
been identified \cite{review1, review2, review3, review4, menon94}.

A related phenomenon is that of aging \cite{kob00, agingexpts}, the observation that a
glassy sample is not in equilibrium and thus its properties
depend on the age of the sample.  Were the system able to
equilibrate at or below $T_{\rm g}$ it would assume the lowest
energy configuration: a crystal.  However upon a temperature
quench to $T<T_{\rm g}$ the system's dynamics are severely slowed
and its structure is frozen in an amorphous configuration.
Slight thermal motions do survive this temperature quench and
slowly allow the system to change configurations albeit in a
non-ergodic fashion.  The time scale for rearrangements grows
with the age of the sample; the older the sample, the longer the time 
needed for a further change.  Similarly to the problem of the
glass transition, a changing time scale for aging dynamics is
suggestive of an underlying changing structural length scale.
One conjecture is that as the glass ages it
optimizes its packing. Specifically the length scale over which
some ``optimal packing'' is achieved might grow during
aging, thus slowing the dynamics of the sample, which in turn
would slow the aging process itself.

Suspensions of colloids have been studied as a model
glass-forming system with great success.
  Colloidal suspensions consist of solid particles in a
liquid, and exhibit a glass transition as their concentration
is increased.  In such systems, the control parameter is
not temperature but packing fraction, $\phi$, and a glass is
obtained once $\phi$ increases above $\phi_{\rm g}\approx0.58$ \cite{pusey86, blaaderen95, snook91, weeks00}.
Previous work with colloids \cite{courtland, cipelletti03} has shed some light on the aging process showing, for example,
that aging is both spatially and temporally heterogeneous.
However a detailed understanding of the microscopic mechanisms
involved is still lacking.

In this work we analyze aging in a colloidal glass from a
geometrical point of view. In particular we focus on the idea
of geometrical frustration \cite{nelson84, nelson02}. In
hard sphere systems the free energy is exclusively determined
by entropy and, in order to minimize the former at a given
temperature, the system has no choice but to maximize
the latter. This is evident already at a packing fraction
of $\phi_{\rm crystallization}\approx .494$ where the system
crystallizes, despite the lack of inter-particle interactions,
sacrificing some configurational entropy to gain vibrational
entropy instead \cite{hoover68}. This is achieved by re-distributing the
available space homogeneously so that all particles have more
local free volume and thus room to move about their average
lattice position.

The most efficient packing of four spheres in three dimensions is to pack them
into a regular tetrahedron.  The effective volume fraction
of four such spheres is $\phi_{\rm tetrahedron}\approx0.78$.
However regular tetrahedra don't tile 3-D space, and thus the
most efficient space filling configuration is a hexagonally
closed packed crystal with $\phi_{\rm hcp}\approx 0.74$.  So while
tetrahedral packing could locally maximize vibrational
entropy, the constraint that the sample fill space induces
crystallization. This {\em frustration} between local and global
packing optimization has been invoked as a possible cause for
the glass transition in simple liquids  \cite{nelson84, nelson02}.

We use fast laser scanning confocal microscopy to observe
micron-sized particles in a colloidal glass in three
dimensions. Specifically we study the static structure of
the glass in terms of tetrahedral packing and analyze the
evolution of some geometrical properties as the sample ages.
While the dynamics change dramatically over the course of our
experiment, we find no change in the structural properties,
as characterized by tetrahedra arrangements.

\section{Experimental Details}

We use poly(methyl methacrylate) (PMMA) particles of diameter
$d=2.36\mu m$ and polydispersity $\sim5\%$. The particles are
sterically stabilized by a thin layer of poly-12-hydroxystearic
acid, dyed with rhodamine 6G and suspended in a mixture of
cyclohexylbromide and decalin (85:15 by weight) that closely
matches both the density and index of refraction of the
particles. This is done to minimize sedimentation effects and to
improve imaging. The solvent's viscosity at 295 K is $\eta=2.25$
mPa$\cdot$s which sets the time required to diffuse a distance
$d$ in the {\em dilute} limit to $\frac{d^{2}}{6D}=\tau_{\rm diff}=23s$
where $D=\frac{k_{\rm B}T}{3\pi \eta d}$.

We use a confocal microscope to acquire three-dimensional
images of a viewing volume of 62 $\mu m$ $\times$ 58 $\mu m$
$\times$ 10 $\mu m$ at a rate of 1 frame every 26 s. Note that
the sample is glassy -- at these densities the colloids move
very slowly.  The particles can be tracked even though the frame rate is
comparable to $\tau_{\rm diff}$.
The viewing volume typically
contains $\sim2500$ particles. We focus at least 60 $\mu m$
away from the cover slip of the sample chamber to avoid wall
effects \cite{kose76, gast86}. We identify particles with a horizontal accuracy
of 0.03 $\mu m$ and a vertical accuracy of 0.05 $\mu m$,
and track them in three dimensions over the course of the
experiment \cite{crocker96, dinsmore01}.

The control parameter for colloidal phase behavior is
the sample volume fraction $\phi$.  While the rhodamine
imparts a slight charge upon the particles, their phase
behavior, $\phi_{\rm freeze}=0.38$ and $\phi_{\rm melt}=0.42$,
is similar to that of hard spheres ($\phi_{\rm freeze}=0.494$
and $\phi_{\rm melt}=0.545$). We observe a glass transition at
$\phi_{\rm g}\approx0.58$, in agreement with what is seen for
hard spheres. We examine samples with volume fractions in the
range  $\phi\approx 0.58-0.62$ and find qualitatively identical
results. These samples form small crystals which nucleate at
the cover slip, but do not form crystals within the bulk of
the sample even after several weeks. Here we report results
based on a sample at $\phi\approx0.62$.

The sample is initialized by vigorous stirring. The subsequent
particle dynamics are reproducible after this stirring,
and depend only on the waiting time $t_{\rm w}$ since the start of
the aging process.  Within a minute of ending the stirring,
transient flows within the sample greatly diminish, and the
particles move slowly enough to be identified and tracked. This
defines our initial time $t_{\rm w}=0$, or age zero, for the
sample.  The results below are not sensitive to variations of
this choice.  During the experiments, no crystallization is
observed within the viewing volume. Note that in many aging
studies, the initial sample is prepared by a temperature quench
(corresponding to a rapid increase of the volume fraction in
our experiments). The initial conditions in our experiments
correspond to a shear-melted sample; the volume fraction
remains constant.

Henceforth all numerical values will be expressed in terms
of particle diameter $d$ and characteristic diffusion time
$\tau_{\rm diff}$.

\section{Results}

\begin{figure}[t]
\centering
\includegraphics[height=5.5cm]{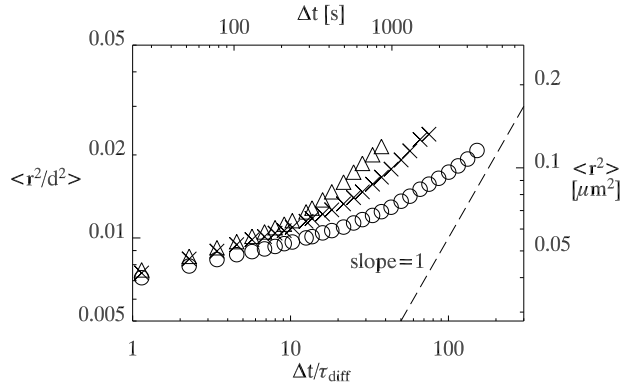}
\caption{Aging mean squared displacement for a
colloidal glass at $\phi\approx0.62$. The three curves
represent three different ages of the sample. $\triangle:
t_{\rm w}=0\tau_{\rm diff}$, $\times:t_{\rm w}=50\tau_{\rm diff}$
and $\bigcirc:t_{\rm w}=150\tau_{\rm diff}$. The dashed line has
a slope of 1 and represents diffusive behavior, not seen in this
glassy sample.}
\label{msd}
\end{figure}

In order to determine whether our sample is aging we take a long
($\sim350\tau_{\rm diff}$) data set and split it in three time
windows as follows: $[0-50\tau_{\rm diff}]$, $[50-150\tau_{\rm
diff}]$ and $[150-350\tau_{\rm diff}]$. Since the sample is not
disturbed during observation this is equivalent to running
three experiments with  $t_{\rm w}=0$, 50, and $150\tau_{\rm
diff}$ from the sample initialization. We then measure the
mean squared displacement for the three data sets averaged
over all particles and over all initial times within a given
window. As can be seen in Fig.~\ref{msd} the data clearly
show a slowing down of the dynamics. For example, as the
sample ages ($t_{\rm w}$ goes from $0$ to $150\tau_{\rm diff}$)
the time particles take to diffuse, say, a distance of $0.02 d^{2}$
goes from approximately $\Delta t\approx30\tau_{\rm diff}$
to $\Delta t\approx90\tau_{\rm diff}$.

\begin{figure}[t]
\centering
\includegraphics[height=5.5cm]{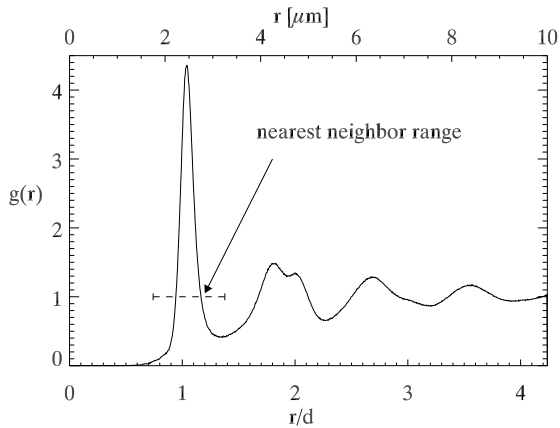}
\caption{Pair correlation function for a sample with $\phi
\approx 0.62$. The dashed segment indicates
the range of interparticle distances used as a definition of
nearest neighbor.
}
\label{gr}
\end{figure}

While the dynamics are clearly changing with the age of the
sample, we wish to consider the static structure as well.
We start by calculating the pair correlation function
$g(r)$. The resulting curve is plotted in Fig.~\ref{gr},
which has been calculated from data averaged over all ages of
the sample; no change in $g(r)$ is seen as the sample ages.
The first peak of $g(r)$ is at $r=1.04d$, slightly displaced
from the ideal hard-sphere position of $r=d$. This is mainly
due to a slight charge induced by the dyeing process, and
may also be due to some uncertainty in the exact value of
the average particle size $d$ which we estimate to be 2\%.
Another effect of this soft potential is that $g(r)$ for $r<d$
does not fall to zero as sharply as it would for perfect
hard spheres.  Figure \ref{gr} also shows the characteristic
double peak around $r=2d$ seen in many glassy systems.

To further characterize the geometry of the colloidal glass we
identify tetrahedra formed by the particles as follows. Two
colloids are labeled as nearest neighbors if their distance is
within the first peak of $g(r)$, namely, if $0.74d>r>1.38d$
as shown in Fig. ~\ref{gr}. The upper bound corresponds to
the first minimum of $g(r)$ while the lower bound is set to
eliminate some nearest neighbor pairs that are excessively
close to each other as a result of occasional errors in
particle tracking. The results presented here are insensitive
to variations in the limits for $r$ and to the use of Delaunay
triangulation as an alternative nearest neighbor finding method.

A tetrahedron is then defined as a quadruplet of mutually
nearest neighbor colloids.  The six inter-particle distances, or
``bond'' lengths, $b_{\rm i}$ with $i=0...5$ are measured.  We then
assign to each tetrahedron a ``looseness'' as measured by the
average bond length $b$ and an ``irregularity'' as measured by
the standard deviation of the six bond lengths  $\sigma_{\rm b}$
and defined by $\sigma_{\rm b}^{2}=\sum^{5}_{\rm i=0}(b-b_{\rm i})^{2}$.
Note that the looseness has the same limits as the nearest
neighbor distance ($b \in [0.74d,1.38d]$) while the irregularity
is bound by zero (for an ideal tetrahedron) and by the maximum
of the standard deviation of six numbers drawn from the interval
$[0.74, 1.38]$, namely $\sigma_{\rm b}\in [0, 0.35d]$.

\begin{figure}[t]
\centering
\includegraphics[height=2.5cm]{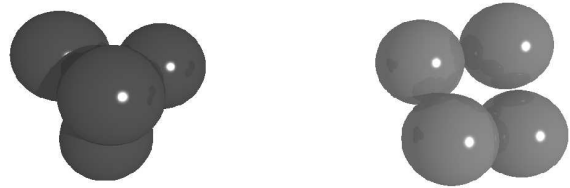}
\caption{Computer generated picture of a very regular
tetrahedron ($\sigma_{\rm b}=0.02d$, left) and a very irregular
one ($\sigma_{\rm b}=0.22d$, right). The spheres are drawn to
scale ($d=2.36\mu m$).}
\label{compare-reg-irreg}
\end{figure}

Figure \ref{compare-reg-irreg} shows a computer rendered
image of a very regular and a very irregular tetrahedron with
$\sigma_{\rm b}=0.02d$ and $\sigma_{\rm b}=0.22d$ respectively. Clearly
our definition of what constitutes a tetrahedron combined
with the range we set for nearest neighbor distances allows
for fairly irregular, quasi planar tetrahedra; though setting
$r_{\rm max}=1.38d<\sqrt{2}d$ avoids the possibility of a perfectly planar
configuration.

\begin{figure}[t]
\centering
\includegraphics[height=5.5cm]{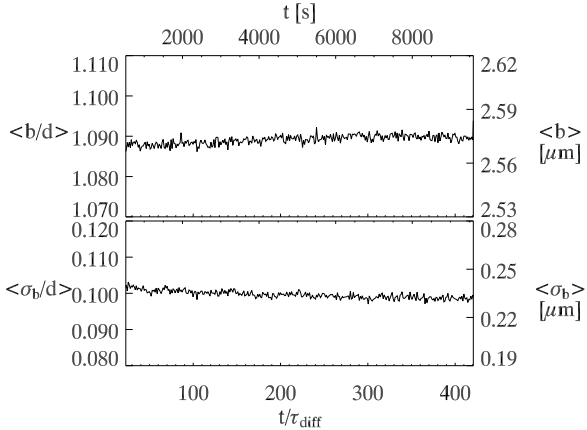}
\caption{Plot of the average tetrahedral ``loosness'' $\langle b\rangle$ (top) and average tetrahedral irregularity $\langle \sigma_{\rm b} \rangle$ (bottom) versus time. The average is taken over all tetrahedra at each time.}
\label{sigma-b-t}
\end{figure}

Since aging is the evolution of the system's properties over
times much longer than $\tau_{\rm diff}$ we analyze these
two static quantities, $b$ and $\sigma_{\rm b}$, as the sample
ages. If aging is a result of thermal fluctuations slowly
trying to resolve the frustration we referred to above,
then one might expect the geometrical characteristics of
the tetrahedra forming the sample to evolve with time. We
therefore tracked the values of $\langle b \rangle$ and
$\langle \sigma_{\rm b} \rangle$ averaged over all tetrahedra in
each frame as the sample ages, and show this in Fig.~\ref{sigma-b-t}.
These two average quantities remain essentially constant over
the duration of the experiment.  $\langle b \rangle$ varies
by less than 0.5\%, reflecting the fact that the packing
fraction $\phi$ remains constant.  One could envisage that
a change in the value of $\langle \sigma_{\rm b} \rangle$ would
allow the sample to find a configuration with lower energy.
For example, a ``better packed'' configuration might have
fewer irregular tetrahedra, as is the case for a crystal.
Such a configuration might allow each particle to have more
local room to move, thus increasing its vibrational entropy.
This however is not the case, as is seen in Fig.~\ref{sigma-b-t},
which shows that $\langle \sigma_{\rm b} \rangle$ decreases by less
than $0.5\%$ throughout the experiment.  While a decrease is
consistent with our argument, this decrease is not reproducible
in other samples, and in fact we find in all cases that $\langle
\sigma_{\rm b} \rangle$ varies by at most $1\%$ with no preference
for increasing, decreasing, or merely fluctuating.

As neither of these average static quantities evolve as
the sample ages, we consider a more complete picture by
calculating the probability distribution functions $P(b/d)$ and
$P(\sigma_{\rm b}/d)$. For each value of the $t_{\rm w}$ the distributions
were calculated by averaging over all tetrahedra and all times
in each time window.

\begin{figure}[b]
\centering
\includegraphics[height=5.5cm]{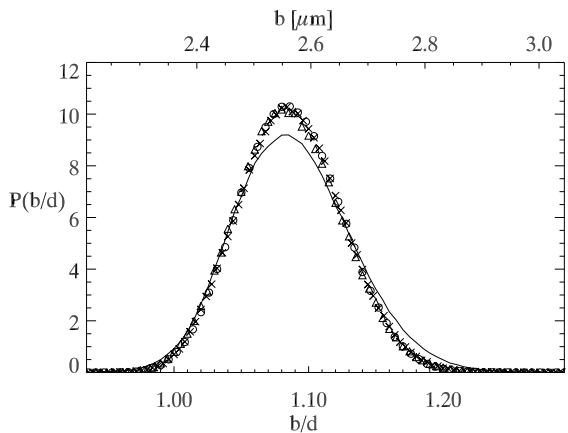}
\caption{The distribution function $P(b/d)$ for tetrahedral bond
lengths (``looseness'') for three ages.
$\triangle: t_{\rm w}=0$, $\times:t_{\rm w}=50 \tau_{\rm diff}$, and
$\bigcirc:t_{\rm w}=150 \tau_{\rm diff}$.  
The solid line is a reference curve computed by considering
random bonds based on $g(r)$, without the constraint that such
bonds form a tetrahedron.
}
\label{Pb}
\end{figure}

The distribution of average bond lengths $P(b/d)$ is shown in
Fig.~\ref{Pb} and is peaked at $b/d \approx \langle b/d \rangle
= 1.09$.  $P(b/d)$ is approximately a Gaussian with a width
of 0.04, although the distribution is slightly asymmetric
with a bias toward larger values of $b$.  Thus compact
tetrahedra ($b/d < 1.05$) are less prevalent, along with
loose tetrahedra ($b/d > 1.13$).  More significantly, the
distribution $P(b/d)$ is unchanged during aging, as is seen
by the excellent agreement between the distributions plotted
with the different symbols. Thus, while both compact and loose
tetrahedra are less likely, their relative prevalence remains
unchanged as the sample ages.  This seems somewhat surprising,
as entropy might favor a more standardized tetrahedron size.
Compact tetrahedra have less room for their particles to move;
the opposite is true for loose tetrahedra.  The system could
maximize its overall vibrational entropy by sharing free
volume more evenly.  However, this is not reflected
in the data, and may indicate that random packing requires
both compact and loose tetrahedra in order to tile space.
Thus, the age of the sample is not detected by $P(b/d)$.

\begin{figure}[t]
\centering
\includegraphics[height=5.5cm]{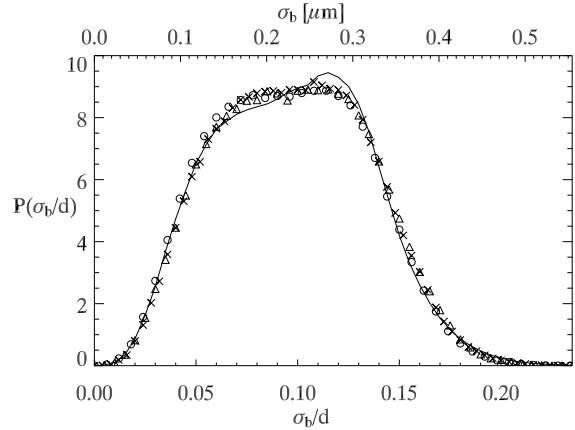}
\caption{The distribution function $P(\sigma_{\rm b}/d)$ for tetrahedral ``irregularity'' for three ages:
$\triangle: t_{\rm w}=0$, $\times:t_{\rm w}=50 \tau_{\rm diff}$, and
$\bigcirc:t_{\rm w}=150 \tau_{\rm diff}$.
The solid line is a reference curve computed by considering
random bonds based on $g(r)$, without the constraint that such
bonds form a tetrahedron.
}
\label{Ps}
\end{figure}

Now consider the distribution of tetrahedral irregularities
$P(\sigma_{\rm b}/d)$ plotted in Fig.~\ref{Ps}. The
most striking feature is a broad maximum around
$\sigma_{\rm b}/b\approx 0.1$. This is congruent with the amorphous
quality of a glassy packing where, unlike in a crystal, one
might expect to find a great variety of tetrahedra, from quite
regular to quite irregular ones. However, it is also interesting
to note that the probability of finding a tetrahedron with
an irregularity greater than 0.2 is effectively zero while
the theoretical maximum for $\sigma_{\rm b}/d$ is $0.35$. Also,
while not very regular, the shape of $P(\sigma_{\rm b}/d)$ is
qualitatively reproducible with other samples above
$\phi_{\rm g}$.  

The striking agreement in the tetrahedral properties between
different ages of the sample suggests that there may be a simple
underlying explanation for these distributions.  One hypothesis
is that the shape of these distributions is dictated by the
statistics of choosing six bond lengths $b_{\rm i}$ at random from the
overall distribution $P(b_{\rm i})$ in an amorphous configuration.
$P(b_i)$ is closely related to $g(r)$ by $P(r) = r^2 g(r)$.  By
randomly picking many sets of six $b_{\rm i}$, without the constraint
that they form a tetrahedron, we can compute reference distributions
$P'(b/d)$ and $P'(\sigma_{\rm b}/d)$.  These are shown as solid lines
in Figs.~\ref{Pb} and \ref{Ps}.  While these are similar to the
experimental data, deviations are clearly seen.  In
Fig.~\ref{Pb}, the solid curve shows a slight bias toward larger
values of $b/d$ as compared with the experimental data.  In
Fig.~\ref{Ps}, the peak of the distributions differs noticeably
between the experimental values and the reference curve.  While
these differences are not extreme, they are outside of the
uncertainty of the data, and thus indicate that tetrahedral
properties do possess additional information beyond just nearest
neighbor bond lengths [which in turn are related to $g(r)$].

\section{Conclusion}

We have examined an aging colloidal glass by looking at
static structural characteristics of tetrahedral packing. We
find that there is no correlation between the age of a glass
and its static structure as evaluated by these quantities.
Nevertheless, our sample's ``awareness'' of its age must be encoded in the
particle positions, and thus in some feature of the packing of
the particles.  We have checked other geometrical quantities
based on tetrahedral structure, such as volume or surface area,
and none of these show aging either.  Tetrahedral properties
appear not to reflect the age and thus a static measure of a
sample's age remains elusive.


\begin{theacknowledgments}
We thank Carrie Nugent and Ted Brzinski for useful discussions.
This work was supported by NASA microgravity fluid physics grant NAG3-2728.
For correspondence contact ERW: weeks@physics.emory.edu.
\end{theacknowledgments}

\bibliographystyle{aipproc}

\end{document}